
%
%
%
%
\input harvmac.tex

\font\smallcapfont=cmr9
\def\sc#1{{\smallcapfont\uppercase{#1}}}
\def\frac#1#2{{\textstyle{#1\over #2}}}
\def\narrow#1{{\leftskip=1.5cm\rightskip=1.5cm\smallskip #1 \smallskip}}

\def\b#1{\kern-0.25pt\vbox{\hrule height 0.2pt\hbox{\vrule
width 0.2pt \kern2pt\vbox{\kern2pt \hbox{#1}\kern2pt}\kern2pt\vrule
width 0.2pt}\hrule height 0.2pt}}
\def\ST#1{\matrix{\vbox{#1}}}
\def\STrow#1{\hbox{#1}\kern-1.35pt}
\def\bv{\b{\phantom{1}}}

\def\tri#1#2#3#4#5#6#7#8#9{\matrix{#4\cr
	#3\quad#5\cr #2~\qquad #6\cr #1\quad #9\quad#8\quad#7\cr}}

\def\arrow#1{{\buildrel#1\over\rightarrow}}
\def\text#1{\quad\hbox{#1}\quad}
\def\gh{\hat{g}}
\def\gb{\bar{g}}
\def\lab{{\bar{\lambda}}}
\def\la{\lambda}
\def\mub{{\bar{\mu}}}
\def\nub{{\bar{\nu}}}
\def\L{\langle}
\def\R{\rangle}
\def\m{{\rm mult}}
\def\lra{\leftrightarrow}
\def\su{\widehat{su}}

\newcount\eqnum
\eqnum=0
\def\eq{\global\advance\meqno by1 \eqno(\secsym\the\meqno)}
\def\eqlabel#1{\eq {\xdef#1{\secsym\the\meqno}}}

\newwrite\refs
\def\startreferences{
 \immediate\openout\refs=references
 \immediate\write\refs{\baselineskip=14pt \parindent=16pt \parskip=2pt}
}
\startreferences

\refno=0
\def\aref#1{\global\advance\refno by1
 \immediate\write\refs{\noexpand\item{\the\refno.}#1\hfil\par}}
\def\ref#1{\aref{#1}\the\refno}
\def\refname#1{\xdef#1{\the\refno}}

\def\pl#1#2#3{Phys. Lett. {\bf B#1#2#3}}
\def\np#1#2#3{Nucl. Phys. {\bf B#1#2#3}}

\newcount\exno
\exno=0
\def\Ex{\global\advance\exno by1{\noindent\sl Example \the\exno:
\nobreak\par\nobreak}}

\parskip=6pt
\Title{\vbox{\baselineskip12pt
\hbox{LAVAL-PHY-20/92}\hbox{LETH-PHY-2/92}\hbox{hepth@xxx/9203004}}}
{\vbox {\centerline{Can fusion coefficients be calculated}
\bigskip
\centerline{from the depth rule ?}
}}

\centerline{A.N. Kirillov$^\flat$, P. Mathieu$^\natural$\foot{Work supported
by NSERC (Canada) and
FCAR (Qu\'ebec).}, D. S\'en\'echal$^\natural$\foot{Work supported by NSERC
(Canada).} and M.A. Walton$^{\sharp 2}$} \vskip.2in
\centerline
{$^\flat$ \it RIMS, Kyoto University, Kyoto 606, Japan}
\smallskip\centerline{$^\natural$ \it D\'epartement de Physique, Universit\'e
Laval, Qu\'ebec, Canada G1K 7P4}
\smallskip\centerline{$^\sharp$ \it Physics Department, University of
Lethbridge,
Lethbridge (Alberta) Canada T1K 3M4}
\vskip .2in
\centerline{\bf Abstract}
\bigskip
\noindent
The depth rule is a level truncation of tensor product
coefficients expected to be sufficient for the evaluation of fusion
coefficients. We reformulate the depth rule in a precise way, and show
how, in principle, it can be used to calculate fusion coefficients.  However,
we
argue that the computation of the depth itself, in terms of which the
constraints on tensor product coefficients is formulated, is problematic.
Indeed, the elements of the basis of states convenient for calculating tensor
product coefficients do not have a well-defined depth! We proceed by showing
how
one can calculate the depth in an `approximate' way and derive accurate lower
bounds for the minimum level at which a coupling appears. It turns out that
this method yields exact results for $\widehat{su}(3)$ and constitutes an
efficient and simple algorithm for computing $\widehat{su}(3)$ fusion
coefficients. \Date{2/92; revised 9/92}

\newsec{Introduction}

By now various methods have been proposed to calculate fusion rules in WZNW
models [\ref{D. Gepner and E. Witten, \np278
(1986) 493.}\refname\GW-\aref{E. Verlinde, \np300
(1988) 389.}\aref{M.A. Walton, \np340 (1990) 77; \pl241 (1990) 365;
V. Ka\v c, {\it Infinite dimensional Lie algebras}, 3rd ed. Cambridge
University Press; F. Goodman and H. Wenzl, Adv. Math. {\bf
82} (1990) 244.}\refname\KW\aref{M. Spiegelglas, \pl245 (1990) 169; P. Furlan,
A. Ganchev and V.B. Petkova, \np343 (1990) 205; J. Fuchs and P. van Driel,
\np346 (1990) 632.}\aref{C.J. Cummins, P. Mathieu and M.A. Walton, \pl254
(1991) 390; L. B\'egin, P. Mathieu and M.A. Walton, J. Phys. A: Math. Gen. {\bf
25} (1992) 135.}\refname\CMW\aref{D. Gepner, Comm. Math Phys.
{\bf 141} (1991) 381}\refname\G\ref {M. Crescimanno, MIT preprint CTP \#2021;
M. Bourdeau, E.J. Mlawer, H. Riggs, and H.J. Schnitzer, Brandeis preprint
BRX-TH-327; D. Gepner and A. Schwimmer, Weizmann preprint
WIS-92/34.}\refname\cres]. The first method to be proposed was the depth rule
of
Gepner and Witten [\GW]. However it has not received much attention since it
did
not appear to be very practical, mainly because of the difficulties involved in
calculating the depth. Therefore the depth rule has not been tested seriously
as
a calculation tool for fusion coefficients. Indeed, except for $\su(2)$, only
low level examples of $\su(3)$ [\GW] and $\hat{E}_8$ [\ref{P. Forgacs, Z.
Horvath, L. Palla and P. Vecsernyes, \np308 (1988) 477.}] have been worked out.
In this work we examine carefully the depth rule as a computational tool for
fusion rules. Section 2 contains a review of the original derivation of the
rule
(see also [\ref{L. Alvarez-Gaum\'e, C. Gomez and G. Sierra, in {\it Physics and
Mathematics of Strings}, edited by L. Brink et al., World Scientific
(1990).}]).

Recall that a fusion rule tells us which primary fields can arise in the OPE
of two given primary fields. More abstractly it defines a product $\times$ as
$$ \la\times\mu = \sum_\nu N^{(k)\ \nu}_{\la\mu}\ \nu $$
where $\la$, $\mu$ and $\nu$ are primary fields. The fusion coefficient
$N^{(k)\ \nu}_{\la\mu}$ gives the number of times $\nu$ appears in the product
$\la\times\mu$.
Notice further that $N^{(k)\ \nu}_{\la\mu} = N^{(k)}_{\la\mu C\nu}$, where
$C\nu$ is the field conjugate to $\nu$.
For WZNW models the spectrum generating algebra is a Ka\v c-Moody algebra
$\gh$ at level $k$ and the primary fields are in one-to-one correspondence
with integrable representations of $\gh$ [\GW,\ref{V.G. Knizhnik and
A. Zamolodchikov, \np247 (1984) 83.}\refname\KZ].

\newsec{The depth rule of Gepner and Witten.}

Let us start by writing explicitly the commutation relations for the untwisted
Ka\v c-Moody algebra $\gh$ at level $k$ [\ref{P. Goddard and D. Olive,
Int. J. Mod. Phys. {\bf A1} (1986) 303.}]:
$$\eqalign{
[H^i_m,H^j_n] &= km\delta^{ij} \delta_{n+m,0}\cr
[H^i_m,E^\alpha_n] &= \alpha^i E^\alpha_{n+m} \cr
[E^\alpha_m,E^\beta_m] &= {2\over\alpha^2}\left(\alpha\cdot H_{n+m}+
km\delta_{n+m,0}\right)\text{if}\alpha=-\beta \cr
&=\epsilon(\alpha,\beta)~E^{\alpha+\beta}_{n+m}\text{if}
\alpha+\beta\text{is a root}\cr
&=0\text{otherwise}\cr}\eqlabel\algebra $$
Here $\alpha$ and $\beta$ denote roots and $\epsilon(\alpha,\beta)$ is a
cocycle.
To each node (i.e. each simple root) of the extended Dynkin diagram one can
associate at least two distinct $su(2)$ subalgebras. In particular for the
zeroth root, whose finite part is $-\theta$ ($\theta$ being the longest root),
the explicit forms of the generators $J^\pm$ and $J_3$ of the two $su(2)$
subalgebras we will use are
$$\eqalign{
(i)\qquad &E^\theta_0~~,~~ E^{-\theta}_0~~,~~\theta\cdot H_0 =
[E^\theta_0,E^{-\theta}_0]\cr
(ii)\qquad &E^{-\theta}_1~~,~~ E^\theta_{-1}~~,~~k-\theta\cdot H_0 =
[E^{-\theta}_1,E^\theta_{-1}]\cr}\eqlabel\gen $$
The finite Lie algebra $\gb$ associated with $\gh$ is recovered from (\algebra)
by setting $m=n=0$ (this corresponds to the homogeneous gradation).
Hence the first $su(2)$ subalgebra in (\gen) is generated by finite Lie algebra
operators while the second one has an intrinsic affine structure.
Now let $\la$ denote an affine weight of positive level $k$, whose finite
part $\lab$ is an integrable highest weight.
There is a one-to-one correspondence between such affine weights and WZNW
primary fields, and we denote them by the same symbol.
To every field $\la$ we associate a state $\la(0)|0\R=|\la\R$.
The degenerate nature of the WZNW primary fields translates into the following
constraints for the singular vectors [\GW]:
$$(E^{-\theta}_0)^{(\la,\theta)+1}|\la\R = 0 \eqlabel\singA $$
$$(E^\theta_{-1})^{k-(\la,\theta)+1}|\la\R = 0\eqlabel\singB $$
Of course there are also singular vectors with $\theta$ replaced by an
arbitrary simple root $\alpha$.

These singular vectors lead to constraints on three-point functions. The
constraints arising from the first type of singular vectors (\singA) being
associated with finite Lie algebra generators, they are taken into account in
ordinary tensor product coefficients of finite Lie algebras [\ref{D.P.
Zelobenko, {\it Compact Lie Groups and their Representations}, Am. Math Soc.,
Providence, 1973; K.R. Parthasarathy, R. Ranga Rao, and V.S. Varadarajan, Ann.
Math., {\bf 85} (1967) 383; S. Kumar, Invent. Math. {\bf 93} (1988) 117.}]. On
the other hand, the other singular vectors yield new constraints on the
ordinary
tensor product coefficients which depend explicitly on the level $k$. Since
$(\la,\theta)\geq(\la,\alpha)$ for any root $\alpha$ of $\bar g,$ the maximal
constraint is obtained by considering the longest root, i.e., precisely from
(\singB). Let us then derive the explicit form of this constraint. For this
consider the three-point function  $$\L
((E^\theta_{-1})^p\la)(z)\mu''(z_1)\nu''(z_2)\R \eqlabel\threeP $$ where
$\mu''$
is associated to a weight (also at level $k$) whose finite part $\mub''$
belongs
to the representation with highest weight $\mub$ (similarly for $\nu''$). This
vanishes if $p>k-(\la,\theta)$ as a result of (\singB). By considering the
$E^\theta_n(z)$ as the modes of the field $E^\theta(\zeta)$ in a Laurent
expansion about $z$, i.e. $$ E^\theta(\zeta) = \sum_n E^\theta_n(z)
(\zeta-z)^{-n-1} $$ one can write $$ E^\theta_{-1}(z) = \oint_z {d\zeta\over
\zeta-z} E^\theta(\zeta) $$ (Factors of $(2\pi i)^{-1}$ in front of contour
integrals are everywhere implicit).
The three-point function (\threeP) can thus be transformed into
$$\oint_z{d\zeta_1\over \zeta_1-z}\dots\oint_z{d\zeta_p\over \zeta_p-z}
\L E^\theta(\zeta_1)\dots
E^\theta(\zeta_p)\la(z)\mu''(z_1)\nu''(z_2)\R$$
Let us now deform the
integration contour of every variable $\zeta_i$ such that it circles around the
singular points $z_1$ and $z_2$.
Expanding the fields $E^{\theta}(\zeta_i)$ in terms of their modes and
performing the integrations (in which only the mode $m=0$ contributes) one
finally gets
$$\sum_{l=0}^p {p!\over l!(p-l)!}{1\over (z-z_1)^l}{1\over (z-z_2)^{p-l}}
\L\la(z)[(E^{\theta}_0)^l\mu''](z_1)
[(E^{\theta}_0)^{p-l}\nu''](z_2)\R $$
As already mentioned, this vanishes if $p>k-(\la,\theta)$.
Since all the terms in the sum are independent, they must all vanish
separately, i.e.
$$\L\la(z)[(E^{\theta}_0)^l\mu''](z_1)
[(E^{\theta}_0)^{p-l}\nu''](z_2)\R = 0\text{if} p>k-(\la,\theta)
\eqlabel\constrA$$
This holds for any value $0\leq l\leq p$.

We conclude that if
$$\eqalign{
(E_0^\theta)^{l_1} \mu'' \equiv \mu' \not= 0\cr
(E_0^\theta)^{l_2} \nu'' \equiv \nu' \not= 0\cr}\eq $$
for $l_1+l_2=p>k-(\lambda,\theta)$, the nontrivial three-point function
$\L \lambda\mu'\nu'\R$ must vanish. Here $\mu'$ (reps. $\nu'$) is characterized
by the fact that one can act $l_1$ (reps. $l_2$) times on it with
$E_0^{-\theta}$ without leaving the representation, i.e., one obtains
a non-zero field $\mu''$ (resp. $\nu''$).

Let us now define the depth of the field $\mu'$ as [\GW]
$$ d_{\mu'} = {\rm max}(l)\text{such that} (E^{-\theta}_0)^l\mu'
\not=0\eqlabel\depth $$
and similarly for $\nu'$ (we will come back to this definition later
and make it more precise). Since $l_1+l_2 < d_{\mu'}+d_{\nu'}$ we conclude
that
$$ \L\la(z)\mu'(z_1)\nu'(z_2)\R=0 \text{if} k < d_{\mu'} + d_{\nu'} +
(\la,\theta)\eqlabel\derul $$ This is the explicit form of the constraints we
were looking for. Of course, irrespective of the value of $k$, it is always
true that $$ \L\la(z)\mu'(z_1)\nu'(z_2)\R=0 \text{if}
{\bar N}_{\lab\mub\nub}=0$$
where ${\bar N}_{\lab\mub\nub}$ denotes the coefficient associated with
the finite tensor product $\lab\otimes\mub\otimes\nub\supset 0$. Since
$$ {\bar N}_{\lab\mub\nub} = {{\bar N}_{\lab\mub}}^{~C\nub} $$
where $C\nub$ is the conjugate of $\nub$, ${\bar N}_{\lab\mub\nub}$ is equal
to the number of times $C\nub$ appears in the tensor product
$\lab\otimes\mub$. (Recall that a necessary condition for
${\bar N}_{\lab\mub\nub}\not=0$ is the existence of a $\mub'$ (resp. $\nub'$)
in the highest weight representation $\mub$ (resp. $\nub$) such that
$\lab+\mub'+\nub'=0$).
We summarize these results in the following {\it depth rule} [\GW]:

\narrow{\noindent\it
The coupling $\L\la\mu'\nu'\R$ vanishes if either
$k<d_{\mu'}+d_{\nu'} +(\la,\theta)$ or ${\bar N}_{\lab\mub\nub}=0$.}

Before proceeding to a detailed analysis of the depth rule, let us emphasize
the following remarkable fact. As is clear from (\depth), the depth $d_{\mu'}$
is a property of the finite part of the weight $\mu'$ (i.e. it involves a
generator of the finite Lie algebra).
Hence, starting from a singular vector involving the action of a generator not
in the finite Lie algebra, namely $E^\theta_{-1}$, which amounts to a
constraint on the affine weight, one ends up with a condition whose only
residual affine characteristic is the level. Thus in applying this rule, all
computations are done with finite weights.

Notice further that the constraint (\singB) is void when $k\rightarrow\infty$.
In that case we are left with purely finite Lie algebra constraints, so that
$$\lim_{k\rightarrow\infty}{N^{(k)}_{\la\mu\nu} = {\bar
N}_{\lab\mub\nub}} \eqlabel\fupo$$
\newsec{Deepening the depth rule}

\subsec{States vs. weights.}

The derivation of the depth rule, reviewed in section 2, depends crucially
upon considering correlation functions, which themselves involve states. We
have pointed out that WZNW primary fields are in one-to-one correspondence
with affine weights $\la$ of positive level $k$ whose finite parts are highest
weights $\lab$. These in turn are in one-to-one correspondence with states
$|\la\R$. This one-to-one correspondence between weights and states no
longer holds if $\la'$ has a finite part $\lab'$ which is not a highest
weight, but rather a degenerate weight with multiplicity $\m(\lab')>1$.
To this weight correspond $\m(\lab')$ states $|\la'_{(i)}\R$,
$i=1,\dots,\m(\lab')$. This clarification is crucial since the depth is a
property of states, not of weights, and it is the states which characterize a
given coupling. The expression (\depth) for the depth should then be sharpened:
$$ d_{\mu_{(i)}'} = {\rm max}(l)\text{such that}
(E^{-\theta}_0)^l|\mu_{(i)}'\R\not=0 \eqlabel\depthBB $$
Similarly the notation $\L \la\mu'\nu'\R$ should now stand for the
correlation $\L \la\mu_{(j)}'\nu_{(\ell)}'\R$ where the doublet $(j,\ell)$ can
take ${\bar
N}_{\lab\mub\nub}$ values.

An aesthetic drawback of the original formulation of the depth
rule is the lack of symmetry between the three states. This can be remedied
by considering, instead of a three-point function, the following four-point
function
$$
\L((E^\theta_{-1})^pI)(0)\la''_{(i)}(z_1)\mu''_{(j)}(z_2)\nu''_{(\ell)}(z_3)\R
$$ which vanishes when $p\geq k+1$. Here $I$ denotes the identity field,
associated with a null finite weight. Following the steps of the previous
section, we obtain that the three-point function $
\L\la'_{(i)}(z_1)\mu'_{(j)}(z_2)\nu'_{(\ell)}(z_3)\R $ vanishes if: $$ k <
d_{\mu'_{(j)}} + d_{\nu'_{(\ell)}} + d_{\la'_{(i)}}\eqlabel\kminB $$ where
$\mub'+\nub'+\lab'=0$. Notice that this reduces to (\derul) when $\lab'=\lab$,
since $d_\la=(\la,\theta)$). Eq. (\kminB) is a
symmetrized version of the depth constraint, which makes manifest the symmetry
of the fusion coefficient $N^{(k)}_{\mu\nu\la}$ under the interchange of any
two
labels.

However, for practical calculations, it is preferable to fix two of the
states, that is to set $\la'_{(i)} = \la$ and $\nu'_{(\ell)} = -C\nu$ and
consider then the three-point function $$
\L\la(z_1)\mu'_{(i)}(z_2)(-C\nu)(z_3)\R, ~~~~i =1,...,{\bar
N}_{\lab\mub\nub}\eqlabel\coup$$
Here we stress that the $\mub'_{(i)}$'s are understood to be the states which,
when coupled to $\lab$ and $-C\nub$, give the scalar representation.
The advantage of (\coup) is that only one out of the three weights is
neither a highest nor a lowest weight, which greatly simplifies the
calculations.
Since the depth of
 $-C\nub$ is obviously zero, (\coup) vanishes whenever
$$k < d_{\mu'_{(i)}
} + (\la,\theta)\ \ .\eqlabel\kzeroi$$

\subsec{Threshold level $k_0^{(i)}$ and the reformulation of the depth rule}

When $k\rightarrow
 \infty$, there are no affine constraints, so that
$\L\la\mu'_{(i)}(-C\nu)\R\not= 0$.  This actually holds true for all values
of $k$ sufficiently large that one does not `hit' the singular vector
(\singB).  For
$k < d_{\mu'_{(i)}
} + (\la,\theta)$, the coupling must vanish as a result of the depth
constraint.
As a result, the smallest value of
the level at which the coupling will be nonzero is
$$ k_0^{(i)}\ \equiv\ d_{\mu'_{(i)}}+(\la,\theta)\ \ .\eqlabel\kminBB $$

For each of the ${\bar
N}_{\lab\mub\nub}$ couplings associated to the tensor product
$\lab\otimes\mub\otimes\nub\supset 0$, there is then a threshold level,
$k_0^{(i)}$, such that for $k\geq k_0^{(i)}$ the coupling is non zero, while
for
$k< k_0^{(i)}$
 it vanishes.
Accordingly, the fusion coefficients are completely
determined by the tensor product coefficients and by the possible values of
$k_0^{(i)}$ given above.  Suppose now that the values of $k_0^{(i)}$ are
ordered such that $k_0^{(i)}\leq k_0^{(i+1)}$.  We can then reformulate the
depth rule in a precise way as:
$$\eqalign{N^{(k)}_{\la\mu\nu}
&= {\rm max}(i)~~ {\rm such~that}~~ k\geq k_0^{(i)} ~~{\rm
and}~~ \bar{N}_{\lab\mub\nub}\not=0\cr
&= 0~~ {\rm if}~~ k< k_0^{(1)} ~~{\rm
or}~~ \bar{N}_{\lab\mub\nub}=0\cr}\eqlabel\newdep $$
with $k_0^{(i)}$ defined in (\kminBB) in terms of the depth.

It remains to see how to calculate the depth.  But first we introduce an
auxiliary concept, the depth charge.

\subsec{Introducing the ``depth charge'' $a_\theta$.}

Suppose we have a basis of states such that $E^{-\theta}_0$ applied to an
element of this basis either vanishes or gives another single element. That is,
one never gets linear combinations of basis elements by applying
$E^{-\theta}_0$. Then the depth
is essentially the maximum number of times one can act on a given state with
$E^{-\theta}_0$ without leaving the representation. If we denote by $|jm\R$ the
projection of this state on the $su(2)$ subalgebra associated to $\theta$
($J^2=j(j+1)$ and $m$ is the $J_3$ eigenvalue) then the depth is simply $j+m$.
We will write $$ a_\theta = 2j \eqlabel\athetaJ$$ and call $a_\theta$ the {\it
depth charge}. For a state $|\mu'_{(i)}\R$, $m$ is the eigenvalue of
$\theta\cdot H_0$, and is given by $$ (\mu',\theta) = 2m $$
Therefore, the depth can be reexpressed as
$$ 2d_{\mu_{(i)}'} = a_\theta(\mu_{(i)}') +
(\mu',\theta)\eqlabel\depthC $$
The calculation of $k_0^{(i)}$ then boils down to that of $a_\theta$.
In terms of $a_\theta$, (\kminBB) reads
$$ 2k_0^{(i)} =
a_\theta(\mu'_{(i)})+(\la+\nu,\theta)\eqlabel\kminD$$
where we used $(\mu',\theta)=(C\nu -\lambda,\theta) = (\nu-\la,\theta)$.

\subsec{Calculation of the depth.}

In order to test
the depth rule, one needs to know how to calculate the depth charge. For this
one must answer the following three questions:
\item{1.} How can we describe states explicitly?
\item{2.} How can we characterize the three states appearing
in a three-point function?
\item{3.} Given a state, how do we calculate the depth charge $a_\theta$?

These questions are looked at in the next section.
{}From now on the discussion will be restricted to $su(N)$ for simplicity.
However the results presented here can be generalized to other algebras.

\newsec{The depth machinery}

\subsec{A basis for states: standard tableaux.}

An $su(N)$ integrable highest weight
$\lab=\sum_{i=1}^{N-1}\la_i\omega^i = (\la_1,\dots,\la_{N-1})$, where
$\omega^i$ is the $i$th fundamental weight, has a natural representation in
terms of a Young tableau with $\la_1+\dots+\la_{N-1}$ boxes on the first row,
$\la_2+\dots+\la_{N-1}$ boxes on the second row, up to $\la_{N-1}$ boxes
on the $(N-1)$th row.
Standard tableaux (also called semi-standard
tableaux [\ref{E. Date, M. Jimbo and T. Miwa, in
{\it Physics and Mathematics of Strings}, edited by L.
Brink et al., World Scientific (1990).}]) are Young
tableaux in which each box is numbered according to the following rule. Let
$c_{i,j}$ be the number appearing in the box on the $i$th row and the $j$th
column. These numbers must satisfy the following constraints:
$$ 1\leq c_{i,j}\leq N \quad,\quad c_{i,j} \leq c_{i,j+1}\quad,\quad c_{i,j} <
c_{i+1,j} $$
In words, the numbers are non-decreasing from left to right and
strictly increasing from top to bottom [\ref{I. Macdonald, {\it Symmetric
functions and Hall polynomials}, Clarendon Press,
Oxford 1979.}\refname\Tableaux].

The allowed standard tableaux of shape $\lab$ are in one-to-one correspondence
with the states in the highest weight representation $\lab$. Weights $\lab'$
with multiplicity $\m(\lab')$ are associated with $\m(\lab')$ different
standard tableaux. {\it Thus standard tableaux provide a convenient basis for
states.}

Notice that the weight of a standard tableau is obtained by adding the weights
of all its boxes. The latter are easily obtained from the fundamental
representations: $\omega^k$ is associated with a single column
of $k$ boxes containing the numbers $1,2,\dots,k$.

\Ex
For $su(3)$ the standard tableaux of the
highest weight states of the fundamental representations are
$$ (1,0)\lra\ST{\STrow{\b1}}\quad,\quad (0,1)\lra\ST{\STrow{\b1}\STrow{\b2}}
\quad,\quad (0,0)\lra\ST{\STrow{\b1}\STrow{\b2}\STrow{\b3}}$$
from which we conclude that $(-1,1)\lra\ST{\STrow{\b2}}$ and
$(0,-1)\lra\ST{\STrow{\b3}}$.
The adjoint representation $(1,1)$ contains 8 standard tableaux:
$$
\matrix{\ST{\STrow{\b1\b1}\STrow{\b2}}\cr (1,1)}\quad
\matrix{\ST{\STrow{\b1\b2}\STrow{\b2}}\cr (-1,2)}\quad
\matrix{\ST{\STrow{\b1\b3}\STrow{\b2}}\cr (0,0)}\quad
\matrix{\ST{\STrow{\b1\b1}\STrow{\b3}}\cr (2,-1)}\quad
\matrix{\ST{\STrow{\b1\b2}\STrow{\b3}}\cr (0,0)}\quad
\matrix{\ST{\STrow{\b1\b3}\STrow{\b3}}\cr (1,-2)}\quad
\matrix{\ST{\STrow{\b2\b2}\STrow{\b3}}\cr (-2,1)}\quad
\matrix{\ST{\STrow{\b2\b3}\STrow{\b3}}\cr (-1,-1)}\quad
$$
Notice that to the doubly degenerate weight $(0,0)$ one associates two distinct
standard tableaux, i.e. two distinct states.

In the following we will use an equivalent representation of this basis,
the so-called Gelfand-Tsetlin patterns [\ref{I.M. Gel'fand and M.L. Tsetlin,
Dokl. Akad. Nauk. SSSR {\bf 71 } (1950) 8, 825, 1071.}\refname\GTref].
To a given standard tableau
we associate the following triangular array of numbers:
$$\matrix{
\beta_1^{(N)}~\beta_2^{(N)}~\cdots\cdots~\beta_N^{(N)}\cr
\beta_1^{(N-1)}~\cdots~\beta_{N-1}^{(N-1)}\cr
\cdots\cdots\cr
\beta_1^{(2)}~\beta_2^{(2)}\cr
\beta_1^{(1)}\cr}\eqlabel\GT $$
such that $\beta_i^{(j)}$ is the number of boxes containing numbers less or
equal to $j$ in the $i$th row of the standard tableau .

\Ex
The following standard tableau and Gelfand-Tsetlin (GT) pattern
corresponding to the weight $(-2,1,0)$ in the representation
$(1,2,1)$ of $su(4)$ are equivalent:
$$\ST{\STrow{\b1\b2\b2\b4}\STrow{\b2\b3\b4}\STrow{\b3}} \quad\cong\quad
\matrix{4~3~1~0\cr 3~2~1\cr 3~1\cr 1\cr} $$

\subsec{Identifying states in tensor products.}

Given a triple product involving a highest weight state $\lab$ and a lowest
weight state $-C\nub$, what are the standard tableaux of shape $\mub$ and
weight $\mub'=C\nub-\lab$ that can appear in the product?
To answer this question we first write down the $\m(\mub')$ standard tableaux
of weight $\mub'$. Among these, the ${\bar N}_{\lab\mub\nub}$ standard
tableaux contributing to the product $\lab\otimes\mub\otimes\nub$ are those
which, when inserted column by column from right to left into the Young tableau
of $\lab$, still give a regular tableau at every step.
The insertion is made by adding to the $i$th row of the Young tableau the
boxes of the standard tableau marked by $i$ [\ref{P. Littelmann, J. of
Algebra {\bf 130} (1990) 328; J. Weyman, Contemporary Mathematics {\bf 88}
(1989) 177; T. Nakashima, RIMS-783 (1991).}].

\Ex
Consider the product $(1,1)\otimes(2,1)\supset(1,0)$.
We are looking for standard tableaux of shape $(2,1)$ and weight $(0,-1)$.
There are two of them
$$ \ST{\STrow{\b1\b3\b3}\STrow{\b2}}\text{and}
\ST{\STrow{\b1\b2\b3}\STrow{\b3}}$$
Adding the first standard tableau to $\ST{\STrow{\bv\bv}\STrow{\bv}}$
(the Young tableau of shape $(1,1)$) column by column from right to left,
one ends with a non-regular tableau after 2 steps: the third row has two
boxes and the second row only one. Thus the first standard tableau does not
contribute to the product, while the second one does (as is easily checked).

We now know how to characterize the states in triple products involving one
highest and one lowest weight state.
In general
it is also possible to describe all the states in a given triple product, but
this requires the introduction of further concepts which are not central in
our argument, and consequently they are introduced in the appendix only.
\subsec{Calculation of the depth charge for the standard tableaux.}

A GT pattern indicates the chain of embeddings
$u(2)\subset u(3)\subset\dots\subset u(N)$ where the $su(2)\subset u(2)$
subalgebra is defined with respect to the root $\alpha_1$, the $su(3)\subset
u(3)$ subalgebra is defined with respect to the roots $\alpha_1$ and
$\alpha_2$,
etc [\GTref]. As a result the spin $j$ of a given state with respect to the
$su(2)$ subalgebra associated to $\alpha_1$
can be read off directly from the GT pattern. Denoting by
$a_{\alpha_1}$ twice this spin, one has
$a_{\alpha_1}=\beta^{(2)}_1-\beta^{(2)}_2$, where the $\beta$'s are defined
just after Eq.(\GT).
Thus the properties of the states with respect to the $su(2)$ subalgebra
associated with $\alpha_1$ are easily extracted.

On the other hand the action of the operators $E^{\pm\theta}_0$ on standard
tableaux or GT patterns is not so simple.
Acting with $E^{\pm\theta}_0$ on a standard tableau generically yields
a linear combination of standard tableaux with the same weight.
Hence a GT pattern does not have a well-defined depth charge. A first
consequence of this observation is that only lower and upper bounds for
$a_\theta$ can be calculated for states whose weights have multiplicity $>1$.
These bounds are the minimal and maximal values that $a_\theta$ can take for
all the states corresponding to a given weight (not just for the states
contributing to the tensor product).\foot{Notice however that for a coupling
$\L\la\mu'\nu'\R$ such that $\nu'=-C\nu$ and $\m(\mub')={\bar
N}_{\lab\mub\nub}$ we can say better. Indeed if we know all possible values of
$a_\theta$ which corresponds to the weights, these values can be used in
eq.(\kminD) to calculate the possible values of $k_0$.} As far as $k_0$ is
concerned only the lower bound is relevant, but does not turn out to be very
useful.

Thus one faces a problem with the depth rule approach for calculating
fusion rules. In this scheme fusion coefficients are truncated tensor product
coefficients. However the basis convenient for the calculation of tensor
product coefficients (namely, standard tableaux) is not convenient when time
comes to evaluate the degree of truncation appropriate to a certain level $k$
in
fusion rules.

\subsec{Introducing the crystal depth charge $a^c_\theta$.}

In spite of this situation, we will try to make further progress, motivated
by the following observation. It is well known that fusion rules
for $\widehat{su}(N)$ at level $k$ are
equivalent to restricted tensor products in the quantum version of
$su(N)$ with $q$ given
by [\ref{L. Alvarez-Gaum\'e, C. Gomez and G. Sierra, \np315 (1989) 155;
\pl220 (1989) 142; \np330 (1990) 347. V. Pasquier and H. Saleur, \np330
(1990) 523.}]:
$$ q = \exp\left\{{2\pi i \over k+N}\right\} $$
On the other hand, at $q=0$ the structure of the quantum group simplifies
considerably [\ref{M. Kashiwara, Comm. Math. Phys. {\bf 133} (1990) 249; C.R.
Acad. Sci. Paris, t. 311 s\'erie 1 (1990) 277.}\refname\CB]
(the group `crystalizes').
For our purpose it is remarkable that an explicit realization  of a
crystal base is given by the standard tableaux [\ref{M. Kashiwara and
T. Nakashima, preprint RIMS-786 (1991).}]\refname\CBST.
Furthermore, it is a particularity of $q=0$ that standard tableaux have a
well-defined value of $a_\theta$. This prompts us to use these values of
$a_\theta$ computed for the quantum group at $q=0$, which we denote
$a_\theta^c$ (the $c$ stands for {\it crystal}), in the formula for $k_0$.
Although the logical motivation for making such a naive extrapolation from
$q=0$ to $q=\exp\{2\pi i/(k+N)\}$ may look weak, it turns out that
this procedure leads us to rather accurate results and, for $\su(3)$, exact
results!

One complication did arise, however. There is a one-to-one
correspondence between the three
couplings
$$\L\la\mu'_{(i)}(-C\nu)\R,~~\L\mu\nu'_{(i)}(-C\la)\R,
{}~\L\nu\la'_{(i)}(-C\mu)\R\ \ .$$
(This correspondence can be seen for instance via the Berenstein-Zelevinsky
triangles, to be introduced in section 6.) We found that the values of
$k_0^{(i)}$ calculated in this way differed in general, for the 3 couplings
above.
It was therefore necessary to take the maximum of the values so obtained. We
symbolise this by rewriting (\kminBB) as
$$ k_0^{(i)}\ =\ {\rm max perm}\{d_{\mu'_{(i)}}+(\la,\theta)\}\ \
.\eqlabel\kminMP $$

As already pointed out, standard tableaux or GT patterns naturally single out
the $su(2)$ subalgebra associated with the root $\alpha_1$. Thus,
in order to calculate $a_\theta$ we will make a suitable `rotation' of the
GT pattern such that $a_\theta$ will be the difference of the new entries
sitting on the line second from the bottom.
But since this operation is justified at $q=0$ only, it is $a^c_\theta$
that we obtain in this way.
Unfortunately, in order to define the explicit operation of `rotation' some
more technology must be introduced.

Let us define the operators $t_i$ which act on the GT patterns as
follows [\ref{A.N. Kirillov and A.D. Berenstein, preprint
RIMS-866 (1992).}\refname\KB]: $$ t_i~:~\left\{\eqalign{
\beta^{(k)}_j&\rightarrow\beta^{(k)}_j \text{if} k\not=i\cr
\beta^{(i)}_j&\rightarrow {\rm min}(\beta^{(i+1)}_j,\beta^{(i-1)}_{j-1})
+{\rm max}(\beta^{(i+1)}_{j+1},\beta^{(i-1)}_j)-\beta^{(i)}_j\cr}\right. $$
If $\beta^{(i-1)}_{j-1}$ is absent it is set equal to $\infty$ and similarly
if $\beta^{(i-1)}_j$ is absent it is set equal to zero.
Then let us introduce the operator
$$ \sigma = t_1t_2\dots t_{N-1} $$
for $su(N)$. These operators have been introduced to describe the action of
the Weyl group directly on GT patterns. The basic Weyl reflections $s_i$ (with
respect to $\alpha_i$) are simply [\KB]
$$ s_i = \sigma^{(i-1)} t_1 \sigma^{-(i-1)} $$
This is equivalent to the action defined by Lascoux-Sch\"utzenberger directly
on the standard tableaux [\ref{A. Lascoux and M.-P. Sch\"utzenberger,
Quaderni della Ricerca {\bf 109} (1981) 129.}].

\Ex
Here are various operations on a particular state of weight $(0,1,-1,0)$ in
the highest weight representation $(0,1,1,1)$ of $su(5)$ producing in the end
a state of weight $(0,0,1,-1)$:
$$\ST{\STrow{\b1\b1\b3}\STrow{\b2\b2\b5}\STrow{\b4\b4}\STrow{\b5}}~\cong~
\matrix{3~3~2~1~0\cr3~2~2~0\cr3~2~0\cr2~2\cr2\cr}~\arrow{t_4}~
\matrix{3~3~2~1~0\cr3~3~1~0\cr3~2~0\cr2~2\cr2\cr}~\arrow{t_3}~
\matrix{3~3~2~1~0\cr3~3~1~0\cr3~2~1\cr2~2\cr2\cr}$$
$$~\arrow{t_2}~\matrix{3~3~2~1~0\cr3~3~1~0\cr3~2~1\cr3~1\cr2\cr}~\arrow{t_1}~
\matrix{3~3~2~1~0\cr3~3~1~0\cr3~2~1\cr3~1\cr2\cr}~\cong~
\ST{\STrow{\b1\b1\b2}\STrow{\b2\b3\b4}\STrow{\b3\b5}\STrow{\b5}} $$
This sequence of operations is nothing but the action of $\sigma$.

Now, $\sigma$ turns out to be the desired `rotation' operator.
Indeed, from the results of [\KB] one can show that
$$ E^{-\theta}_0 = \sigma^{-1} E^{\alpha_1}_0\sigma\ \ . $$

\Ex
The action of $\sigma=t_1t_2$ on the adjoint representation of
$su(3)$ is shown on Fig.1.
One sees that this action amounts to a rotation of the peripheral weights by
$120^\circ$ and to an interchange of the two inner states. The $\theta$
direction has been rotated to the $\alpha_1$ direction except for a `sign',
which illustrates the fact that $E^{-\theta}_0$ (a $J_-$ operator) is actually
related to $E^{\alpha_1}_0$ (a $J_+$ operator).

In terms of GT patterns this means that $a^c_\theta$ can be calculated as
follows.
Let us write
$$\tilde{GT} \equiv \sigma(GT) $$
where the entries of $\tilde{GT}$ are denoted $\tilde\beta_j^{(i)}$. Then
$$ a^c_\theta = \tilde\beta_1^{(2)} - \tilde\beta_2^{(2)} $$

\Ex
The following state corresponds to the weight $(0,-1,0)$ in the $su(4)$
highest weight representation $(1,1,1)$:
$$\ST{\STrow{\b1\b3\b4}\STrow{\b2\b4}\STrow{\b3}}~\cong~
\matrix{3~2~1~0\cr2~1~1\cr1~1\cr1\cr}~\arrow{\sigma}~
\matrix{3~2~1~0\cr3~1~0\cr3~0\cr2\cr}~:~a^c_\theta = 3-0=3 $$
Similarly the value of $a^c_\theta$ for the standard tableau
considered in example 4 is 2.

\subsec{A simple rule to calculate $a^c_\theta$ for $su(3)$.}

In this section we derive a simple rule for computing the value of $a^c_\theta$
directly from standard tableaux in the case of $su(3)$.
In this case $\sigma=t_1t_2$, but since $t_1$ affects only the last row of the
GT pattern, it is sufficient to consider only $t_2$. Hence one has
$$\matrix{a~b~c\cr d~e\cr f\cr}~\arrow{t_2}~
\matrix{a~b~c\cr \tilde{d}~\tilde{e}\cr f\cr} $$
where
$$\eqalign{
\tilde{d} &= a + {\rm max}(b,f)-d\cr
\tilde{e} &= {\rm min}(b,f)+c-e\cr}$$
Dropping from the outset columns of three boxes from the standard tableau, one
can set $c=0$ (this amounts to setting the irrelevant $u(1)$ charge of
$u(1)\otimes su(N)\subset u(N)$ to zero). Thus
$$ \eqalign{
a^c_\theta &= a-d+e+{\rm max}(b,f)-{\rm min}(b,f)\cr
&= a-d+e+|b-f|\cr}$$
This is easily checked to be equivalent to
$$ a^c_\theta = \sum_{columns} |\hbox{number of 1's}-
\hbox{number of 3's}|\eqlabel\atheta $$
In other words, the value of $a^c_\theta$ of a given standard tableau  can be
calculated by adding the values of $a^c_\theta$ corresponding to each of its
columns. Furthermore, in each column the value of $a^c_\theta$ is just the
number of 1's minus the number of 3's (in absolute value).

\Ex
$$
a^c_\theta\left(\ST{\STrow{\b1\b1\b2\b2\b2\b3}\STrow{\b2\b3\b3}}\right)~=~3$$

\newsec{Testing the depth rule with $a^c_\theta$.}

We are now in a position to test the depth rule with crystal depth
charge through various examples.

\subsec{The $\su(3)$ case.}

\Ex

Let us first consider a classic example, namely the triple product
$(1,1)\otimes(1,1)\otimes(1,1)$. It is well-known that
$(1,1)\otimes(1,1)=2(1,1)\oplus\dots$.
In fusion rules one copy of $(1,1)$ arises at level 2 while the other
appears at level 3. Indicating the value of $k_0$ by a subscript one
has
$$ (1,1)\otimes(1,1)=(1,1)_2 \oplus(1,1)_3\oplus\dots $$
To the weight $(0,0)$ correspond two standard tableaux of shape $(1,1)$:
$$\ST{\STrow{\b1\b2}\STrow{\b3}}\text{and}\ST{\STrow{\b1\b3}\STrow{\b2}} $$
Their values of $a_\theta^c$ are respectively 0 and 2 (c.f. (\atheta)).
Thus here (\kminD) gives directly $k_0=2,3$, which is the correct result
(permutations are not necessary since the three weights are equal).

Using a computer program we have tested the depth rule through (\kminMP)
with the depth given by (\depthC) and $a_\theta$ calculated at
$q=0$ by (\atheta). We computed all fusion coefficients up to level 10 and
found perfect agreement with the results obtained from the Ka\v c-Walton
formula [\KW]. This led us to suspect that the depth rule
with crystal depth charge is correct for $\su(3)$.
Actually, this can be proved directly by relating our algorithm to
other known algorithms for $\su(3)$.
We will return to this in the next section.

\subsec{An $\su(4)$ counterexample.}

Here we present an example which shows that the depth rule
with crystal depth charge is not correct in general.
The simplest counterexample that we found is the following $\su(4)$ product:

\Ex

$$ (1,2,1)\otimes (1,2,1) = 4(1,2,1)_5\oplus(1,2,1)_6\oplus\dots\eqlabel\exA $$
To the weight $(0,0,0)$ in the representation $(1,2,1)$ correspond seven
standard tableaux, among which five (obtained by the method of section 4.2)
contribute to the product:
$$\ST{\STrow{\b1\b1\b2\b3}\STrow{\b2\b3\b4}\STrow{\b4}}\qquad
\ST{\STrow{\b1\b1\b2\b3}\STrow{\b2\b4\b4}\STrow{\b3}}\qquad
\ST{\STrow{\b1\b1\b2\b4}\STrow{\b2\b3\b3}\STrow{\b4}}\qquad
\ST{\STrow{\b1\b1\b3\b3}\STrow{\b2\b2\b4}\STrow{\b4}}\qquad
\ST{\STrow{\b1\b1\b3\b4}\STrow{\b2\b2\b4}\STrow{\b3}}$$
The corresponding values of $a^c_\theta$, calculated according to the rule
given in section 3.3, are respectively $0~,~2~,~2~,~2~,~4$.
Hence formula (\kminD) yields the following values of $k_0$:
$$ k_0 = 4~,~5~,~5~,~5~,~6 $$
which disagrees with \exA.

\newsec{$\su(3)$: relation with other methods.}

In this section we show that, in the case of $\su(3)$, the depth rule
with crystal depth charge is equivalent to all other known algorithms for
computing $\su(3)$ fusion rules. This equivalence is obtained via the
Berenstein-Zelevinsky (BZ) triangles, introduced in section 6.1.
We then reexpress $k_0$ as given by eq.(\kminD) in terms of the entries of
these triangles. Next we recall that the generating function for $\su(3)$
fusion rules obtained in [\CMW] (and whose correctness was proven by Cummins
[\ref{C.J. Cummins, J. Phys. A: Math. Gen. {\bf 24} (1991) 391.}\refname\Cum])
can be translated into a simple formula for $k_0$.
This in turn can also be reexpressed in terms of the parameters of the BZ
triangles. The resulting formula for $k_0$ is equivalent to that obtained in
[\ref{S. Lu, PhD thesis, MIT (1990). A.N. Kirillov,
unpublished.}\refname\Lu] by other methods. We then show that it is also
equivalent to the formula derived from the depth rule with $a_\theta^c$.

\subsec{Berenstein-Zelevinsky triangles for tensor product coefficients.}

Consider the set of three $su(3)$ highest weights $(\la_1,\la_2)$,
$(\mu_1,\mu_2)$ and $(\nu_1,\nu_2)$ (we give here the Dynkin labels).
Berenstein and Zelevinsky [\ref{A.D. Berenstein and A.Z. Zelevinsky,
Cornell preprint - technical report 90-60 (1990).}] showed that the
number of triangles one can construct according to the following rules:
$$\tri{a_1}{a_2~~}{a_3}{a_4}{a_5}{a_6}{a_7}{a_8}{a_9} \text{such that}
\matrix{a_1+a_2=\la_1\cr a_3+a_4=\la_2\cr a_4+a_5=\mu_1\cr
a_6+a_7=\mu_2\cr a_7+a_8=\nu_1\cr a_9+a_1=\nu_2\cr}\qquad
\matrix{a_2+a_3 = a_6+a_8\cr a_3+a_5 = a_9+a_8\cr a_5+a_6 = a_2+a_9\cr }
\eqlabel\BZ $$
gives the value of ${\bar N}_{\lab\mub\nub}$. Such triangles make manifest
most of the symmetries of the tensor product coefficients.

\Ex

Corresponding to the coupling $(2,2)\otimes(2,2)\otimes(2,2)$,
three BZ triangles can be constructed:
$$ \tri022022022 \qquad,\qquad \tri111111111 \qquad,\qquad\tri200200200 $$
and accordingly the multiplicity of the coupling is  3.

The states involved in a specific coupling can be read off a triangle as
follows. Consider the product $\lab\otimes\mub=C\nub\oplus\dots$  associated
with the BZ triangle (\BZ). The state $|\mub'_{(i)}>$ (of weight $C\nub-\lab$)
in this coupling is described by the GT pattern
$$ \matrix{\mu_1+\mu_2\qquad\quad\mu_2\qquad\qquad0\cr\cr
\mu_1+\mu_2-a_4~\quad\mu_2-a_6\cr\cr\mu_1+\mu_2 -a_4 -a_2\cr}\eqlabel\muGT$$

\Ex
The GT patterns and corresponding standard tableaux (in the representation
$\mub$) associated with the three BZ triangles of the last example are
(in the same order)
$$\matrix{4~2~0\cr4~0\cr2}~\cong~\ST{\STrow{\b1\b1\b2\b2}\STrow{\b3\b3}}
\quad,\quad
\matrix{4~2~0\cr3~1\cr2}~\cong~\ST{\STrow{\b1\b1\b2\b3}\STrow{\b2\b3}}
\quad,\quad
\matrix{4~2~0\cr2~2\cr2}~\cong~\ST{\STrow{\b1\b1\b3\b3}\STrow{\b2\b2}}
$$
\subsec{$k_0$ from BZ triangles.}

Consider the BZ triangle (\BZ) describing the coupling between $\lab$,
$\nub'=-C\nub$ and $\mub'$ given by (\muGT).
The value of $a_\theta^c(\mu')$ can be obtained by acting on the GT pattern
(\muGT) with $\sigma=t_1t_2$ (actually $t_2$ alone is enough in this case) and
by taking the difference of the entries on the second line. This gives
$$ a_\theta^c(\mu') = a_4+a_7 + |a_5-a_2| $$
Therefore one has
$$ a_\theta^c(\mu') + (\nu+\la,\theta) = 2\ {\rm max} (a_4+\nu_1+\nu_2,
a_7+\la_1+\la_2)$$
The permuted versions of the above are evaluated similarly. As a result, one
can rewrite (\kminD) in the form
(There are only 3 permutations here because interchanging $\nu$ and $\lambda$
does not change the value of the crystal depth charge
$a_\theta^c(\mu^\prime).$)
For $a_2\leq a_5\leq a_8,$ this boils down to
$$ k_0\ =\ {\rm max}\{a_7+\lambda_1+\lambda_2, a_4+\nu_1+\nu_2\}\ \ .$$
But $(a_4+\nu_1+\nu_2) - (a_7+\lambda_1+\lambda_2) = a_5 - a_2 \geq 0,$ by
assumption, so we obtain
 $$k_0\ =\
a_4+\nu_1+\nu_2\ \ \ . \eqlabel\kminE$$
It can similarly be shown that if we assume instead $a_2\leq a_8\leq a_5,$ the
formula (\kminE) remains valid, and so holds for $a_2\leq{\rm min}(a_5,a_8).$

\Ex
{}From the above formula one readily computes the values of $k_0$ for the three
triangles of example 10 to be respectively 4, 5 and 6.
 (Notice that for these 3 triangles the inequality $a_2\leq{\rm min}(a_5,a_8)$
is satisfied. For less obliging triangles, however, one simply uses their
symmetry to rotate them until the inequality is obeyed.)
\subsec{$k_0$ from the decomposition of BZ triangles into elementary
couplings.}

Recall that every coupling $\lab\otimes\mub\otimes\nub$ can be decomposed into
a  product of elementary couplings $E_i$ [\ref{J. Patera and R.T. Sharp,
in {\it Lecture Notes in Physics}, vol. 84, Springer Verlag,
New York 1979.}\refname\PS]. For $su(3)$ the 8 elementary couplings are

$$\matrix{E_1=(1,0)(0,1)(0,0)\cr~\cr\tri010001000}\qquad
\matrix{E_2=(1,0)(0,0)(0,1)\cr~\cr\tri100000000}\qquad
\matrix{E_3=(0,0)(1,0)(0,1)\cr~\cr\tri000010001}$$

$$\matrix{E_4=(0,1)(1,0)(0,0)\cr~\cr\tri000100000}\qquad
\matrix{E_5=(0,1)(0,0)(1,0)\cr~\cr\tri001000010}\qquad
\matrix{E_6=(0,0)(0,1)(1,0)\cr~\cr\tri000000100}$$

$$\matrix{E_7=(1,0)(1,0)(1,0)\cr~\cr\tri010010010}\qquad
\matrix{E_8=(0,1)(0,1)(0,1)\cr~\cr\tri001001001}$$

However there is some redundancy in such a decomposition since different
products of elementary couplings can be equivalent. Since the BZ triangle of
a product coupling
is simply the sum of the BZ triangles of the factors, one sees that
$E_1E_3E_5 = E_7E_8$, since they both have the same BZ triangle:
$$E_1E_3E_5 = E_7E_8 ~:~\tri011011011 $$
For $su(3)$ this is the only redundancy [\PS], and to obtain a unique
decomposition for a general coupling one simply has to forbid one of the
products $E_7E_8$ or $E_1E_3E_5$ to appear. This leads directly to the
construction of a generating function for tensor product coefficients.

It was conjectured in [\CMW] that there is at least one choice of forbidden
couplings which allows for the construction in a simple way of a generating
function for fusion coefficients from that of tensor product coefficients.
Given
such a choice the conjecture boils down to a simple relation for the minimum
level of a coupling in terms of the corresponding  minimum level of the
elementary couplings appearing in its decomposition. Writing the decomposition
in the form $$ \lab\otimes\mub\otimes\nub = \prod_i E_i^{b_i} $$
the precise relation is
$$ k_0 = \sum_i b_ie_i\eqlabel\elemA $$
where $e_i$ is the minimum level for the elementary coupling $E_i$.

The generating function for $\su(3)$ fusion coefficients obtained from this
method
has been proven to be equivalent to the combinatorial algorithm of Cummins
[\Cum]. This constitutes a proof of the relation (\elemA) for $\su(3)$.
The correct choice of forbidden coupling turns out to be $E_1E_3E_5$.

Conversely, granting (\elemA), it is simple to find which coupling must be
forbidden.
Indeed, since $E_1E_3E_5=E_7E_8=(1,1)(1,1)(1,1)$, let us consider this specific
fusion rule (c.f. example 8).
We know that this coupling comes in two copies, with $k_0=2,3$.
The corresponding BZ triangles together with their decomposition into
elementary couplings are
$$\matrix{E_2E_4E_6\cr~\cr\tri100100100}\qquad
\matrix{E_1E_3E_5=E_7E_8\cr~\cr\tri011011011}\eqlabel\EEE$$
For $\su(3)$ $e_i=1$ for all $i=1,2,\dots,8$, and therefore $E_1E_3E_5$
must be forbidden in order to obtain the correct result $k_0=2$ for
the second triangle.

Having determined the forbidden coupling appropriate for the description of
fusion rules, one can decompose uniquely any coupling into elementary
couplings and read off $k_0$ from (\elemA). By forbidding $E_1E_3E_5$ it is
straightforward to obtain the general and unique decomposition of a BZ triangle
into a product of the $E_i$'s. For instance, assuming that $a_2\leq{\rm
min}(a_5,a_8)$, one finds the following decomposition for (\BZ):
$$ E_2^{a_1}E_3^{a_5-a_2}
E_4^{a_4}E_5^{a_3-a_6}E_6^{a_7}E_7^{a_2}E_8^{a_6}\quad,\quad
(a_2\leq{\rm min}(a_5,a_8))$$
from which it follows that
$$ k_0 = a_4 + \nu_1+\nu_2\quad,\quad (a_2\leq{\rm min}(a_5,a_8))
\eqlabel\kminF$$
in agreement with (\kminE).\foot{An equivalent result has been obtained in
[\Lu].
However, their approach is different since they started from
the Ka\v c-Walton formula, which relates fusion coefficients to
tensor product coefficients using the action of the affine Weyl
group. Furthermore the expression for $k_0$ given by Lu is
different from the one above. This is because his triangle is
related to ours by a Sch\"utzenberger involution $(t_1t_2t_1)$ of the
associated
standard tableau.  Explicitly, Lu's result is $$k_0 (t_1t_2t_1{\rm BZ}) = {\rm
max} [{\rm min}
(a_3,a_8)+\mu_1+\mu_2,
{\rm min}(a_2,a_6)+\nu_1+\nu_2,{\rm min}(a_5,a_9)+\la_1+\la_2].$$} This
proves that the strong depth rule with $a_\theta^c$ is correct for $\su(3)$.

\Ex
The triangle $E_2E_4E_6$ ($k_0=3$) in (\EEE) is associated with the standard
tableau $\ST{\STrow{\b1\b3}\STrow{\b2}}$
(c.f. the relation between (\BZ) and (\muGT)) with $a_\theta^c=2$ and thus
$k_0=3$, while the triangle $E_7E_8$ ($k_0=2$) is associated with
$\ST{\STrow{\b1\b2}\STrow{\b3}}$, with $a_\theta^c=0$ and thus $k_0=2$.

\newsec{Conclusion.}

We have reformulated in a precise way the depth rule of Gepner
and Witten [\GW], and shown how it can be used in principle to
 calculate fusion coefficients.
However, the cornerstone of the depth rule is the depth itself
and we have argued that its calculation is problematic. More precisely, the
depth rule is a constraint imposed on tensor product coefficients. But the
basis convenient for calculating tensor product coefficients is not
convenient for the calculation of the depth.

To carry on, we compute the depth in a special ($q=0$) limit of the quantum
version of the finite part of the affine algebra under consideration, and
freely
extrapolate the result. Unfortunately, for $\su(N)$ $(N>3)$ it does
not give the correct results. However we stress that counterexamples have been
found only for moderately high levels and multiplicities. This indicates that
the depth rule with crystal depth gives rather accurate lower
bounds\foot{We have no proof that it is actually a lower bound, but the
statement is motivated by the very large number of examples we have worked
out.}
 for the minimum levels at which couplings first appear.
On the other hand, for $\su(3)$ this procedure gives exact results. It
(namely, (\muGT) with $a_\theta$ given by (\atheta)) turns out to be
a rather efficient and simple algorithm for the calculation of fusion
coefficients. A little puzzle remains, however: why does it work for $\su(3)$,
and why is it so accurate in general?

As for $\su(3)$, it seems that the success of the present approach is linked to
the fact that all elementary couplings have level one, which implies that the
decomposition of a coupling into elementary couplings can be done column by
column in Young tableaux. In this vein, the failure of the depth rule with
crystal depth charge for $\su(N>3)$ could be explained by the existence of
elementary couplings with minimum level higher than 1. However, thanks to
syzygies, they can most of the time be eliminated from the decomposition of a
coupling into elementary couplings. This should account for the reasonable
success of the crystalline version of the depth rule. Finally, for
$\su(3)$, we noticed that BZ triangles provided a very convenient ground for
the
analysis of generating functions. We will report elsewhere on the extension of
this approach to $\su(N>3)$.

We have argued that in the absence of an appropriate basis, generic
fusion coefficients cannot be easily computed from the depth rule
(which is based on the consideration of three-point functions).
But we should mention another way by which
they could be extracted solely from the analysis of singular vectors.  Indeed,
among the Kac-Moody singular vectors are those that may be interpreted as mixed
Virasoro-Ka\v c-Moody singular vectors, given the Sugawara construction. These
mixed singular vectors lead to the Knizhnik-Zamolodchikov equation [\KZ].  From
this equation, one can evaluate the four-point functions and extract from these
the three-point functions, which then yield the fusion coefficients.
However, to our knowledge, no such calculations have been performed for
three-point functions such that ${\bar N}_{\lab\mub\nub} > 1$.

On the other hand, the depth rule has been
used in [\G] to compute arbitrary fusion coefficients using a Schubert
type calculus.\foot{It should be stressed that this approach is conceptually
very different from the method of generating functions [\CMW].
Starting from the
fact that any irreducible representation can be obtained from an appropriate
multiple product of fundamental representations, it is shown in [\G] that
this description can be extended to the affine case provided we impose some
level dependent constraints, called syzygies, on the ordinary product
of the fundamental representations.  On the other hand, for the method of
generating functions, the basic variables are the elementary couplings and the
syzygies there refer to relations between products of elementary couplings,
already present for the calculation of tensor products.}
This does not contradict our conclusion since the depth rule is used there only
to calculate fusion coefficients involving at least one fundamental
representation (i.e. the truncated Pieri's formula).  For a fundamental
representation, the depth charge is easily evaluated and it agrees with its
crystal version.

\vskip12pt
\centerline{\sc{acknowledgements}}
A.N.K. and P.M. would like to thank RIMS for its hospitality, and in particular
the organizers of the workshop {\it RIMS 91-Infinite analysis} (during which
part of this work was done) for providing a very stimulating environment
especially favorable to fruitful exchange of technologies. M.W. thanks the CERN
Theory Division for hospitality, and A. Ganchev and C. Cummins for useful
discussions.

\appendix{A}{Tensor products with standard tableaux: words, bumping and the
Robinson-Schensted correspondence.}

In this appendix we describe a more general way of representing states
in a given coupling and of calculating tensor product coefficients.
To a standard tableau one may associate a {\it word}, which is the sequence of
numbers appearing in the tableau, read column after column from bottom to top
and from left to right.

\Ex
To the standard tableau
$$ \ST{\STrow{\b1\b1\b2\b3\b5}\STrow{\b3\b4\b4}\STrow{\b5}} $$
we associate the word $[531414235]$.

On the other hand, given a word one can
reconstruct the corresponding standard tableau by the {\it bumping} method.
One places the leftmost number in a box, the upper left box of the tableau
being constructed. For the next number one
proceeds as follows: If it is greater than or equal to the number already
placed in the first box, it is put in the second box of the first row;
otherwise it takes the place of the first number in the first box, the latter
being {\it bumped} in the first box of the second row. Proceeding in this way
for all numbers of the word and allowing bumping in every row, one easily
reconstructs the corresponding standard tableau. This relation between
standard tableaux and words is known as the Robinson-Schensted correspondence
[\ref{C. Schensted, Can. J. Math. {\bf 13} (1961) 179; D.E. Knuth, Pacific J.
Math. {\bf 34} (1970) 709; D.E. Knuth, {\it The art of computer programming},
vol. III, (sect. 5.1.4) Addison Wesley 1973.}].

\Ex
Here is the step-by-step reconstruction of the tableau
associated with the word $[3121]$:
$$\ST{\STrow{\b3}}\leftarrow [121]~~=~~
\ST{\STrow{\b1}\STrow{\b3}}\leftarrow [21]~~=~~
\ST{\STrow{\b1\b2}\STrow{\b3}}\leftarrow [1]~~=~~
\ST{\STrow{\b1\b1}\STrow{\b2}\STrow{\b3}} $$

(This example illustrates the fact that words are invariant with respect to
some rearrangements in the order of the entries: the words $[3121]$ and
$[3211]$ are manifestly equivalent.)

Any intermediate step in the reconstruction of a standard tableau from a word
can be viewed as the insertion of a word into a standard tableau. The
representation of a standard tableau by a word, followed by the insertion of
this word into another standard tableau defines
a `fusion' procedure for two standard tableaux which exactly reproduces the
Littlewood-Richardson rule for calculating tensor products coefficients.
This construction, due to Thomas [\ref{G.P. Thomas, Adv. in Math.
{\bf 30} (1978) 8.}\refname\Thomas], provided the first combinatorial
proof of the Littlewood-Richardson rule.
More explicitly, in order to obtain all possible $\nub$ which could appear in
the product $\lab\otimes\mub$ one proceeds as follows. Consider all the states
$T(\lab_{(i)}')$ (resp. $T(\mub_{(i)}')$)\foot{In our notation $T(\la)$ is the
standard tableau associated with the state $|\la\R$, and $W(T(\la))$ is the
corresponding word.} in the $\lab$ (resp. $\mub$) representation. By fusing all
possible pairs of standard tableaux like $T(\lab_{(i)}')\leftarrow
W(T(\mub_{(j)}'))$, one fills up all the highest weight representations $\nub$
which appear  in the product $\lab\otimes\mub$. Notice that the symmetry of the
tensor product with respect to the interchange of $\lab$ and $\mub$ is
reflected by the existence of a dual word $W^*$ and a right insertion operation
such that [\Thomas] $$ W^*(T(\lab'))\rightarrow T(\mub')~=~T(\lab')\leftarrow
W(T(\mub')) $$

\Ex
Fusion of the ST
$$ \ST{\STrow{\b1\b2\b3}\STrow{\b2\b4}}\text{and}
\ST{\STrow{\b1\b4}\STrow{\b3}\STrow{\b4}} $$
is done as follows:
$$ \ST{\STrow{\b1\b2\b3}\STrow{\b2\b4}}\leftarrow [4314]~~=~~
\ST{\STrow{\b1\b1\b3\b3\b4}\STrow{\b2\b2\b4}\STrow{\b4}} $$
which shows that the representation $(2,2,1)$ appears in the $su(4)$ tensor
product $(1,2,0)\otimes(1,0,1)$.

The generalization of the Thomas construction to the case of a triple product
$\lab\otimes\mub\otimes\nub$ is straightforward: one simply has a chain of
insertions of the type $[T\leftarrow W(T')]\leftarrow W(T'')$.

\Ex
Let us identify the two triplets of states contributing to the $su(3)$
tensor product $(1,1)\otimes(1,1)\otimes(1,1) \supset (0,0)$.
These are
$$\left\{ \ST{\STrow{\b2\b3}\STrow{\b3}}~,~\ST{\STrow{\b1\b2}\STrow{\b3}}~,~
\ST{\STrow{\b1\b1}\STrow{\b2}}\right\}\text{and}
\left\{ \ST{\STrow{\b2\b3}\STrow{\b3}}~,~\ST{\STrow{\b1\b3}\STrow{\b2}}~,~
\ST{\STrow{\b1\b1}\STrow{\b2}}\right\} $$
Indeed, in the first case one has
$$\ST{\STrow{\b2\b3}\STrow{\b3}}\leftarrow[312]~=~
\ST{\STrow{\b1\b2\b3}\STrow{\b2\b3}\STrow{\b3}}$$
followed by
$$\ST{\STrow{\b1\b2\b3}\STrow{\b2\b3}\STrow{\b3}}\leftarrow[211]~=~
\ST{\STrow{\b1\b1\b1}\STrow{\b2\b2\b2}\STrow{\b3\b3\b3}}~=~(0,0)$$
In this example, associativity is easily verified, i.e.
$$[T\leftarrow W(T')]\leftarrow W(T'') ~=~
T\leftarrow W[T'\leftarrow W(T'')] $$

\Ex
With this method one can redo example 9. In order appropriate to left
insertion, the contributing tableaux are
$$\left\{\ST{\STrow{\b2\b3\b3\b4}\STrow{\b3\b4\b4}\STrow{\b4}}\quad,\quad
X\quad,\quad \ST{\STrow{\b1\b1\b1\b1}\STrow{\b2\b2\b2}\STrow{\b3}}\right\}$$
where $X$ stands for five tableaux given in Ex. 9. Since $a^c_\theta$ for the
two other tableaux above is easily checked to be 4, one reobtains the set
$\{4,5,5,5,6\}$ for $k_0$.
\bigskip \hrule \bigskip \centerline{\sc{references}}

\immediate\closeout\refs \vskip 0.5cm
  \message{References}\input references
\bye

\def\b#1{\kern-0.4pt\vbox{\hrule\hbox{\vrule\kern2pt\vbox{\kern2pt \hbox{#1}
\kern2pt}\kern2pt\vrule}\hrule}}
\def\ST#1{\matrix{\vbox{#1}}}
\def\STrow#1{\hbox{#1}\kern-1.2pt}

\def\arrow#1{{\buildrel#1\over\rightarrow}}

\tolerance=1000
\overfullrule=0pt
\nopagenumbers
\hsize=9.15in
\vglue3cm
$$\matrix{
\matrix{\ST{\STrow{\b1\b2}\STrow{\b2}}\cr\bullet}\kern 3cm
\matrix{\ST{\STrow{\b1\b1}\STrow{\b2}}\cr\bullet}\cr\noalign{\vskip2cm}
\ST{\STrow{\b2\b2}\STrow{\b3}}\bullet\kern 3cm
\matrix{\ST{\STrow{\b1\b2}\STrow{\b3}}\cr\bigodot\cr\noalign{\vskip0.15cm}
\ST{\STrow{\b1\b3}\STrow{\b2}}}\kern 3cm \bullet
\ST{\STrow{\b1\b1}\STrow{\b3}}\cr\noalign{\vskip2cm}
\matrix{\bullet\cr\noalign{\vskip0.2cm}\ST{\STrow{\b2\b3}\STrow{\b3}}}\kern 3cm
\matrix{\bullet\cr\noalign{\vskip0.2cm}\ST{\STrow{\b1\b3}\STrow{\b3}}}\cr
}
\kern 2cm\matrix{\sigma\cr\longrightarrow}\kern 2cm
\matrix{
\matrix{\ST{\STrow{\b1\b1}\STrow{\b3}}\cr\bullet}\kern 3cm
\matrix{\ST{\STrow{\b1\b3}\STrow{\b3}}\cr\bullet}\cr\noalign{\vskip2cm}
\ST{\STrow{\b1\b1}\STrow{\b2}}\bullet\kern 3cm
\matrix{\ST{\STrow{\b1\b3}\STrow{\b2}}\cr\bigodot\cr\noalign{\vskip0.15cm}
\ST{\STrow{\b1\b2}\STrow{\b3}}}\kern 3cm \bullet
\ST{\STrow{\b2\b3}\STrow{\b3}}\cr\noalign{\vskip2cm}
\matrix{\bullet\cr\noalign{\vskip0.2cm}\ST{\STrow{\b1\b2}\STrow{\b2}}}\kern 3cm
\matrix{\bullet\cr\noalign{\vskip0.2cm}\ST{\STrow{\b2\b2}\STrow{\b3}}}\cr
}$$
\vglue2cm
\centerline{Fig.1: Action of $\sigma$ on the highest weight representation
$(1,1)$ of $su(3)$.}

\bye